\documentstyle[12pt]{article}
\def\la{\mathrel{\mathpalette\fun <}}

\def\beq{\begin{equation}}
\def\eeq{\end{equation}}
\def\fun#1#2{\lower3.6pt\vbox{\baselineskip0pt\lineskip.9pt
\ialign{$\mathsurround=0pt#1\hfil##\hfil$\crcr#2\crcr\sim\crcr}}}

\begin{document}

\title{Ground State for the Quark Mass Hierarchy and Mixings}
\author{I.T.Dyatlov\\
Petersburg Nuclear Physics Institute\\
Gatchina, St.Petersburg 188350, Russia\\
e-mail: dyatlov@lnpi.spb.su}
\maketitle

In the  last few years the Nambu-Jona-Lasinio (NJL) model \cite{1}
for dynamical chiral violation was widely used to study the possible
influence of high energy phenomena on the formation of the standard
model. At small distances  $(M^{-1}\ll M_w^{-1})$, strong NJL forces
were assumed to couple only with the heaviest quarks \cite{2}, in
which case electroweak symmetry breaking and fermion mass generation
were involved via a top-\-quark condensate. The method \cite{3}
suggested a new way to relate high and low energies and obtain
testable predictions.

However, few attempts are reported to have been made to involve  low
masses and quark mixing \cite{4}. The initial separation of the
top sector only complicates the understanding.

In the present paper we will extend and develop the small distance
part of the scheme suggested in \cite{2,3}. For the purpose we will
extend the NJL equation into the approximation next to the leading
$N_c$ one $(N_c\gg1$ is the number of colors). At first we will
 show that under certain conditions an NJL system symmetric in $n$
flavors may spontaneously choose the heaviest state out of $n$
initially equal massless quarks. The resulting ground state of one
massive $m_1$ and $(n-1)$ massless quarks is the very one to develop
the mass hierarchy and mixings.
Secondly, we will determine conditions in which hierarchy and
mixing can appear. Here the both are considered as small distance
phenomena. Therefore they are to be reflected in the gap equation for
induced masses by flavor dependent NJL coupling constants.
The ground state being asymmetric,
it is sufficient to have a tiny violation of the original flavor
symmetry to have both phenomena generated at the level required. The
asymmetric part of the NJL coupling must meet tough restrictions
$(\la m_1^2/M^2$ or $N^{-1}_c)$ to yield exactly the mass hierarchy.
The same conditions strongly suppress neutral flavor changings.

The delicate problem of the scheme \cite{2,3} is the phenomenological
necessity of fine tuning. It is required that the theory should
closely approach the critical point if the heaviest fermion mass
$m_1\ll M$. Now $m_1/M$ will not determine fine-tuning conditions.
They can be restricted by the small but constant factor $N^{-1}_c$.

The interaction we will consider is similar to that in \cite{5,6}:
\beq
V=\lambda_{ii'}\left(\bar \Psi^c_{Li} \Psi^c_{Ri'}\right)
\left(\bar \Psi^{c'}_{Ri'}\Psi^{c'}_{Li}\right), \qquad
\lambda_{ii'}\sim M^{-2}.
\eeq
Chiral quarks have $n$ flavors (generations) $i,i'$ and $N_c$ colors
$c,c'$. In the general case $\lambda_{ii'} \neq \lambda_{i'i}$. The
theory has the symmetry $SU(N_c)\times[U_L(1)]^n \times[U_R(1)]^n.$
Therefore mixing between flavors may only be spontaneous.
One may believe that Eq.(1) is not an arbitrary chosen example, but
that at low energy $(E<M)$ it could simulate quite a realistic high
energy $(E>M)$ situation \cite{5,7}.

We intend to use two successive $N_c$ approximation in the gap
equation instead of the leading approximation commonly used up to
now. The gap equation for the mass matrix $\Sigma_{ii'}$ is (Fig.1):
\begin{eqnarray}
\Sigma_{ii'}&=&\frac{\beta_{ii'}}{M^2}\int\frac{d^4p}{\pi^2i} \frac14
Sp\,G_{ii'}(p,\Sigma)f \nonumber \\
&+&\frac1{2N_c}\sum_{i_1i_1'}\int\frac{d^4q}{\pi^2i}
G_{i_1i'_1}(-q,\Sigma)
B^{i'i'_1}_{ii_1}(q^2)\frac{\beta_{i'i'_1}}{M^2}, \\
\beta_{ii'}&=&\frac{\lambda_{ii'}M^2 N_c}{8\pi^2} = {\rm const,  }\;\;
\;\mbox{  when  }\;\; N_c\to \infty;  \nonumber
\end{eqnarray}
 $f$ is the cut-off function. We shall use the simplest form: $f
=\vartheta(M^2-|p^2|)$. The cut-\-off form does not influence
qualitative results in case we deal with divergences contained in
simple loops. The sought-for quark propagator is taken in the form:
\beq
G^{-1}_{ii'}(p,\Sigma)=\Sigma_{ii'}-\hat p\delta_{ii'}.
\eeq
Therefore, in Eq.(2) the external momentum $P$ (Fig.1b) is assumed
equal to zero.

In Eq.(2), the quark-antiquark amplitude $N^{-1}_c B\beta/M^2$ can be
calculated as a plait of simple loops (Fig.1b). This basic statement
follows from the analysis of possible multiloop contributions. When
Eq.(2) is valid the general formula for $B$ is
\beq
B\sim\frac{M^2}{m^2_1}\left[1+\frac1{N_c}f_1\left(\frac{m^2_1}{M^2}
\right)+\frac1{N^2_c}f_2\left(\frac{m^2_1}{M^2}\right)+\cdots\right]
\eeq
at $m^2_1\sim q^2$ and $m^2_1\ll M^2$. We do not take into account
the powers $\beta^k$, as $\beta=$const. The first term represents a
sum of simple loops.

Some danger could have appeared if $N^{-1}_c$ was compensated by the
large $M^2/m^2_1$ factor. But such contributions do not exist.
 Then, the utmost we could obtain is
$f_1\sim f_2\sim\ldots\sim0(1)$. However, a part of these
contributions becomes important in the $N^{-1}_c$ order, Eq.(2).
Their role will be discussed below.

The sum of simple loops can be written in a matrix form:
\beq
N^{-1}_c B\beta=N^{-1}_c(1-A)^{-1}\beta, \qquad \beta=\left(
\begin{array}{cc} 0 & \beta_{ii'} \\ \beta_{ii'} &0
\end{array}\right),
\eeq
where $A,B,\beta$ are matrices in the chirality and flavor spaces:
\begin{eqnarray}
A^{\alpha\beta}_{ii_1,i'i'_1}(q)&=&\int\frac{d^4p}{\pi^2i}\frac12 Sp
\left[\frac{\beta^{\alpha\delta}_{ii_1}}{M^2}O_\delta G_{ii'}
(p,\Sigma)O_\beta G_{i_1i'_1}(p-q,\Sigma)\right]f, \nonumber \\
O_{\pm} &=& \frac12 (1\pm\gamma_5).
\end{eqnarray}

The coupling constants we are interested in are almost independent of
flavor indices (see Eq.(20)):
\beq
\lambda_{ii'}=\lambda_0+\delta\lambda_{ii'},\qquad
|\delta\lambda_{ii'}|\ll \lambda_0.
\eeq
At first we will consider Eqs.(2)--(6) at the symmetric
$U_L(n)\times U_R(n)$ interaction (1), i.e. $\delta\lambda_{ii'}=0$.
In this case the unknown matrix $\Sigma_{ii'}$ can be taken in a
diagonal form:  $i$ and $i'$ bases are arbitrary. Solutions we are
interested in vary in the number $n'$ of massive states at equal
$m_1$ masses, $n'\le n$. Flavor indices are preserved in $B$, all
expressions depend only on the chiralities $\alpha,\beta$. Any
element $(i=i'$ and $i_1=i'_1)$ of the matrices $(1-A)$ and $B$ can
be written as \beq \delta_{\alpha\beta}-A_{\alpha\beta}
=A^{(1)}\delta_{\alpha\beta} +A^{(2)}, \qquad
B_{\alpha\beta}=B^{(1)}\delta_{\alpha\beta} +B^{(2)}.
\eeq The
equation $(1-A)B=1$ gives us the part of $B$ participating in Eq.(2)
\beq
B^{(2)}=-\frac{A^{(2)}}{[A^{(1)}+2A^{(2)}]A^{(1)}}.
\eeq

The denominators, Eq.(9), represent propagators of both the bound
massive scalar and goldstone states (after symmetry breaking). These
denominators are equal to
\begin{eqnarray}
A^{(1)}+2A^{(2)}&=&\frac1{M^2}\left[\Gamma\left(\frac{m^2_1}{M^2}
\right)M^2+\frac12\beta(4m^2_1-q^2)I(m^2_1,q^2)\right], \nonumber\\
A^{(1)}&=&\frac1{M^2}\left[\Gamma\left(\frac{m^2_1}{M^2}
\right)M^2+\frac12\beta(-q^2)I(m^2_1,q^2)\right], \\
I(m^2_1,q^2) & = & \int\frac{d^4p}{\pi^2i}\frac1{(m^2_1-p^2)
[m^2_1-(p-q)^2]}f, \;\;\; \; \beta=\frac{\lambda_0 M^2
N_c}{8\pi^2}.\nonumber
\end{eqnarray}
The function $\Gamma(m^2_1/M^2)$  takes the form:
\beq
\Gamma\left(\frac{m^2_1}{M^2}\right)=1-\beta\left(1-\frac{m^2_1}{M^2}
\ln \frac{M^2}{m^2_1}\right)+O\left(\frac1{N_c}\right),
\eeq
it represents the expression for "the goldstone mass" $(\Gamma M^2)$.

The simple loop (Eq.(6)) $(\delta\lambda_{ii}=0)$ contributes the two
first terms to Eq.(11). They strictly coincide with the gap equation
at $N_c\to\infty$ (the Fig.1a contribution). Preservation of
$N^{-1}_c$ terms in the gap equation (2) requires us to take into
account $N^{-1}_c$ terms of the same order in Eqs. (10)--(11). They
must appear from the terms $N_c^{-1}f_1$, $N_c^{-2}f_2,\ldots$ of
Eq.(4). It is impossible to calculate terms with overlapping square
divergences in the unrenormalizable model (1), but there is no need
to do that. In the denominators (Eq.(10)), the constant $\sim
N^{-1}_c$ terms contribute only to the function $\Gamma M^2$
representing the goldstone mass. All other possible terms will be
smaller:  $N_c^{-1}(m^2_1/M^2), N^{-1}_c(q^2/M^2)$.
Therefore, after symmetry breaking we should simply take
$\Gamma=0$.  Then, the remainder of the Eq.(4) terms $\sim N^{-1}_c,
N^{-2}_c,\ldots,$ may be neglected in the gap equation (2). In this
way we are consistently determining the NJL gap equation in the
approximation next to the leading $N_c$ one.

Thus, at $\delta\lambda_{ii}=0$ the second term in rhs (2) is written
as:  \begin{eqnarray}
\frac{n'}{N_cm_1}F(m^2_1)&=&-\frac{2n'}{N_c}\int\frac{d^4q}{\pi^2i}\;
\frac{m^2_1}{m^2_1-q^2}\frac1{(4m^2_1-q^2)(-q^2)I(m^2_1,q^2)}
\nonumber \\ &\to & \frac{-2\sqrt{2}\,n'}{\ln(M^2/m^2_1)N_c}.
\end{eqnarray}

The integral is calculated at $\ln M^2/m^2_1\gg1$. Due to Eq.(9), we
have $F<0$.

In the symmetric case the gap equation is
\beq
1-\beta^{-1}=\frac{m^2_1}{M^2}\ln\frac{M^2}{m^2_1}+\frac{n'}{\beta
N_c}\left|\frac{F(m^2_1)}{m_1}\right|.
\eeq
The $m^2_1>0$ solution exists when
\beq
\beta>1.
\eeq
At $N_c\to\infty$ Eqs.(13),(14) become one loop conditions well known
in NJL models $(\beta=$const).

The energy shifts of massive Dirac cellars \cite{1} enable us to
compare vacuum energies of different solutions $m_1(n')$. We have
\beq
E^{n'}_n[m_1(n')]-E_n(0)=-2n'm^2_1(n')0(\Lambda^2)N_cV,
\eeq
$0(\Lambda^2)$ is the square divergent integral. At $N_c\to\infty$ a
stable solution corresponds to $n'=n$. There exists a region of
parameters $\beta$ and $N_c$ where $n'=1$ state will have the minimal
energy. This region is easy to determine at
$M^2/m^2_1\gg\ln(M^2 m^2_1)\gg1$. In this case the first term in the
rhs (13) may be neglected to obtain the root:
\beq
\frac{m^2_1(n')}{M^2}\simeq
\exp\left[-\frac{2\sqrt{2}\,n'}{(\beta-1)N_c}\right]\ll1.
\eeq
The solution of one massive flavor will be stable if
$(\beta-1)N_c\ll2\sqrt{2}/\ln2.$

This problem can also be solved by substitution of $\frac1\lambda
\phi^2-\phi(\bar \psi\psi)$ for $\lambda(\bar \psi\psi)^2$, eq.(1),
via integration over auxiliary scalar fields $\phi$ \cite{3,6,8}.

It should be noted that different $n'$ solutions exhibit the
following symmetry breakings:
\beq
SU_L(n)\times SU_R(n)\to SU_L(n-n')\times SU_R(n-n')\times SU_V(n').
\eeq

Let us assume $n'=1$ and consider Eq.(2) again. Our aim is to adjust
such $\delta\beta_{ii'}$ properties that would necessarily produce
the mass hierarchy and mixing in all quark generations: $m_1\gg
m_2\gg m_3,\ldots$
$(\delta\beta_{ii'}=\delta\lambda_{ii'}M^2N_c/8\pi^2)$.

Well-known (see \cite{9} and refs. therein) is that the hierarchical
structure can be explained if lower masses appear as radiative
corrections. This mechanism implies that   proper mass matrices
for up and down quarks $\Sigma^{U,D}_{ii'}$ should be represented as
perturbation series in some generation dependent interaction
\cite{7}. Therefore, we will also expand the gap equation (2),
propagators and $B$ amplitudes $(\delta B=-B_0\delta AB_0)$ with
regard of Eq.(7) and the formula:
\beq
 \Sigma_{ii'}=\Sigma_0+\delta\Sigma_{ii'}, \qquad\;
|\delta\Sigma_{ii'}| \ll \Sigma_0 =\frac{m_1}n.
\eeq
In Eq.(18) we seek $\Sigma_{ii'}$ in the general bases of $L,R$
systems, since such bases appear to be more appropriate in case
$\delta\beta_{ii'}$ is unknown. After having eliminated the
non-\-perturbative gap equation (13), we obtain the following form of
the $\delta\Sigma_{ii'}$ system:
$$ \delta\Sigma_{ii'}\left[1-\beta^{-1}-\frac{m^2_1}{M^2}\ln
\frac{M^2}{m^2_1} +\frac1{N_c\beta}f^{(1)}\right]+\frac{\Delta
(\delta\Sigma)_{ii'}}{m^3_1}\left[\frac{m^2_1}{M^2}
\ln\frac{M^2}{m^2_1}+\frac1{N_c\beta}f^{(2)}\right] $$
\begin{eqnarray}
&+&\frac1{N_c\beta}f^{(3)}\left[\sum_i\frac{\delta\Sigma_{ii'}}{
m_1}+\sum_{i'}\frac{\delta\Sigma_{ii'}}{m_1}\right]=-
\frac{\delta\beta_{ii'}}\beta;  \\
\Delta(\delta\Sigma)_{ii'}
&=&\delta_{\delta\Sigma}\left\{\Sigma_{ii'}Sp\Sigma^T\Sigma-(\Sigma
\Sigma^T\Sigma)_{ii'}\right\}, \nonumber
\end{eqnarray}
where $f^{(i)}$ are the functions of $\ln M^2/m^2_1$.
 In order to obtain a normal solution, $m_2\ll m_1$, it is
necessary to have a very small asymmetric coupling constant
$\delta\beta_{ii'}$:
\beq
\delta\beta_{ii'}=0\left(\frac{m^2_1}{M^2},\frac1{N_c}\right)
\delta\tilde \beta_{ii'}, \qquad\; |\delta\tilde \beta_{ii'}| \le
\sqrt{\frac{m_2}{m_1}}
\eeq
(if $\delta\beta_{ii'}$ depends on one index, either $i$ or $i'$,
$m_2/m_1\sim|\delta\beta_i|^2$ or $|\delta\beta_{i'}|^2$). The
quantity $m_2$ and mixing properties will be obtained by
diagonalization of Eq.(18), provided $\delta\Sigma_{ii'}$ is a
solution of Eq.(19). At $N_c\to\infty$ Eq.(19) has no solutions.

If considered outside the model frame, the restriction (20) may turn
out important. The non-\-perturbative part of $B$, Eq.(9), preserves
flavor, therefore all flavor changings may occur via small
non-\-diagonal constants $\sim N_c^{-1}\delta\beta$ (20). Thus,
neutral transitions induced by a new high energy physics could be
strongly suppressed.

Due to the factor $0(m^2_1/M^2, N^{-1}_c)$ the series for
$\beta_{ii'}$ (i.e. for $\lambda_{ii'}$, Eq.(7)) and, hence, the
series for $\Sigma_{ii'}$ (which in reality must be induced by
$\beta$) cannot be usual perturbation expansions. A necessary
situation may occur when strong flavor independent forces give rise
to a critical phenomenon and only small field components can
distinguish flavors \cite{7}. The expansion then takes place in the
ratios between small flavor dependent and strong independent fields.
In this case the factor $\sim0(m^2_1/M^2,N^{-1}_c)$ can coincide with
these ratios.
Thus, it becomes apparent that different fields and forces take
part in quark spectrum phenomena.

The articles \cite{7} contain the detailed analysis of this
quasi-\-perturbative problem. The resulting point is that practically
all qualitative features of the quark mass hierarchy and weak
mixings will be reproduced if small flavor dependent fields are of a
chiral vector type. The extended publication will present Eq.(19)
solution at length.

The author is grateful to Ya.I.Azimov,
M.I.Eides, V.Yu.Petrov, and A.V.Yung for clarifying discussions. This
work was supported by the Russian Fundamental Investigation Fund
(project No.93-02-3145).

Figure caption

{\bf Fig.1.} Equation for the fermion self-energy.

\end{document}